\documentclass[useAMS,usenatbib]{mn2e}
\usepackage{graphicx}
\usepackage{lscape}
\usepackage{ulem}

\def\kms{km~s$^{-1}$}

\def\omgb{$\Omega_b$}
\def\avgnh{$<n({\rm H})>$}

\title[The Galactic Deuterium Abundance and Dust Depletion]
{The Galactic Deuterium Abundance and Dust Depletion: Insights
From an Expanded Ti/H Sample\thanks{Based on 
observations made with ESO Telescopes at the Paranal Observatories under 
programme ID 076.C-0503(A)}}
\author[S. L. Ellison, J. X. Prochaska \& S. Lopez]
{Sara L. Ellison$^1$\thanks{Email: sarae@uvic.ca},
Jason X. Prochaska$^2$, Sebastian Lopez$^3$
\\
$^1$Dept. Physics \& Astronomy, University of Victoria, 3800 Finnerty Rd, 
Victoria, BC, V8P 1A1, Canada\\
$^2$Dept. of Astronomy and Astrophysics, UCO/Lick Observatory,
University of California, 1156 High Street, Santa Cruz, CA 95064, USA\\
$^3$Departamento de Astronom\'ia, Universidad de Chile, Casilla
     36-D, Santiago, Chile\\
}

\begin{document}
\maketitle

\begin{abstract}

The primordial abundance of deuterium (D/H) yields a measure of
the density of baryons in the universe and is an important complement
to determinations from cosmic microwave background (CMB) experiments.  
Indeed, the current small sample of high redshift D/H measurements
from quasar absorption line studies are in excellent agreement
with CMB-derived values.  Conversely, absorption line
measurements of the 
\textit{Galactic} D/H ratio in almost 50 stellar sightlines show
a puzzlingly large scatter outside the Local Bubble which
is difficult to explain simply by astration from the primordial
value.  The currently favoured
explanation for these large variations
is that D is differentially depleted relative to H in some
parts of the local interstellar medium (ISM).  Here, we test this scenario
by studying the correlation between D/H and the abundance of
titanium, one of the most refractory elements readily observed in the ISM. 
Previous work by Prochaska, Tripp \& Howk (2005) found tentative
evidence for a correlation between Ti/H and D/H based on seven sightlines.
Here we almost triple the number
of previous Ti measurements and include several sightlines with very
high or low D/H that are critical for quantifying any correlations with
D/H.   With our larger sample, we confirm a correlation between
Ti/H and D/H at the 97\% confidence level.  However, 
the magnitude of this dependence is difficult to reconcile
with a simple model of dust depletion for two reasons.
First, contrary to what is expected from local depletion rates,
the gradient of the highly refractory Ti is much shallower
than that observed for Fe and Si.  Second, we do not observe the
established tight, steep correlation between [Ti/H] and the mean volume
density of hydrogen.  Therefore, whilst dust remains a plausible
explanation for the local D/H variations, the abundances of
at least some of the refractory elements do not provide unanimous
support for this scenario.  We also argue that the correlations of
[Si/H], [Fe/H], and [Ti/H] with D/H are inconsistent with a simple infall
model of low metallicity gas with approximately solar abundances as
the dominant cause for D variations.

\end{abstract}


\section{Introduction}

The abundances of the light elements (deuterium, helium and lithium)
form one of the cornerstones of Big Bang nucleosynthesis theory and can
all be used as baryometers, yielding the density of baryons in
the universe as a fraction of the closure density, \omgb.  
Although the light element abundances are actually a direct
probe of the baryon-to-photon ratio, $\eta$, this can be translated
to \omgb\ using the (now well-known) Cosmic Microwave Background (CMB) 
temperature.  Deuterium is by far the most sensitive of the light elements
to $\eta$, and thus the baryon density, and determining the primordial value of
D/H (the deuterium to hydrogen abundance) has been the focus of intense
research efforts over the last three decades (e.g. York \& Rogerson
1976; Vidal-Madjar et al. 1977; Laurent, Vidal-Madjar \& York 1979;
Pettini \& Bowen 2001; O'Meara et al. 2001; Hebrard et al. 2002; Moos et 
al. 2002;
Sonneborn et al. 2002; Kirkman et al. 2003; Sembach et al. 2004; Wood
et al. 2004; Oliveira et al. 2006).  These efforts have made
a two-pronged attack, by measuring D/H abundances both in the Milky Way 
and at high redshifts.   Although D is not produced in any significant 
quantities after Big Bang nucleosynthesis, it is easily astrated in stars,
so measurements at high redshifts could potentially yield the
primordial abundance more readily than observations in the local universe.
High redshift measurements can be achieved with observations 
of high HI column density absorption systems in the lines of sight 
towards QSOs, such as the damped Lyman alpha
systems (DLAs) and Lyman limit systems (LLSs). However, since the DI Lyman $\alpha$
line is only separated from that of HI by $\sim$ 80 km/s, only those
rare absorbers
with very simple velocity structure yield reliable results.
Thus, despite a decade of high redshift work, only six robust 
measurements of D/H at high redshift currently exist (O'Meara et al. 2001;
Pettini \& Bowen 2001; Kirkman et al. 2003; O'Meara et al.\ 2006).
Nonetheless, these measurements paint a relatively coherent
picture and yield an \omgb\ (inferred from a mean D/H = (2.8$\pm$0.4)
$\times 10^{-5}$)\footnote{The error on the D/H abundance from
QSO absorption lines is estimated from a jackknife analysis of the 
weighted means of six robust measurements.} that is in solid agreement
(i.e. within the 1 $\sigma$ errors) with that determined from CMB
experiments such as WMAP (see Pettini 2006 for a review).

Measurements of D/H in the Milky Way complement high redshift
observations by providing an order of magnitude more sightlines
in which abundances can be measured (see Linsky et al. 2006 for
the most recent compilation).  Although some astration
has probably occurred, the highest Galactic D/H ratios set a lower
limit to the primordial abundance.  Moreover, modelling
the amount of astration inferred from sub-primordial abundances
provides clues to the rate at which material has been processed
in stars -- a vital ingredient in chemical evolution models.
For these reasons, one of the principle drivers of
the Far-Ultraviolet Spectroscopic Explorer (FUSE) was to
compile a large database of Galactic D/H measurements:
`The main science goal of the FUSE mission has been to obtain accurate
measurements of (D/H)$_{\rm gas}$ for many sightlines in the Milky 
Way Galaxy and beyond in order to measure (D/H)$_{\rm prim}$ and to
obtain constraints on Galactic chemical evolution' (Linsky et al. 2006).  
However, despite
a massive investment of telescope resources from both FUSE and
other satellite facilities, and measurements of
D/H in almost 50 lines of sight, the picture of
Galactic deuterium abundances remains puzzling.
On the one hand, inside the Local Bubble ($<$ 100 pc from
the Sun) the D/H value seems roughly constant at 1.5 $\times 10^{-5}$,
approximately 60\% of the primordial value.  
However, beyond this bound, there is a large scatter in the D/H ratios
which can not be easily explained by astration 
(Jenkins et al.\ 1999; Wood et al.\ 2004; Draine, 2006).
As pointed out by Draine (2006), there is an observed factor of five range
in D/H for clouds within 100 pc.  It is difficult
to reconcile such disparate abundances, on such small spatial scales,
within a chemical evolution model, particularly since mixing
should be efficient on these scales (e.g. Romano et al. 2006).

At the present time, the most favoured explanation for
the variation in Galactic D/H ratios is the depletion of deuterium
from the gas phase onto grains.  Deuterium depletion was
first suggested over 20 years ago by Jura (1982) and then explored
theoretically by Tielens (1983).  The most recent theoretical
work has been done by Draine (2004, 2006) who suggested that
deuterated polycyclic aromatic hydrocarbons (PAHs) 
in cold dust could be sufficient sinks for
deuterium to explain the scatter in Galactic measurements.
Empirical support for D depletion
comes from the composition of interstellar grains collected in
the upper terrestrial atmosphere which can be enriched by D
relative to the usual meteoritic composition (Keller, Messenger \& Bradley
2000).  Whilst the quantity of deuterium was not sufficient to
explain the Galactic ISM abundances, the presence of dust phase
D does act as a `proof of concept'.

One empirical test of the depletion scenario is to look for
correlations between D/H and the abundances of other refractory
elements.  This test was first done by Prochaska, Tripp \&
Howk (2005) using far-blue optical lines of Ti~II measured in
Keck/HIRES spectra for seven Galactic sightlines.  Prochaska et al.
(2005) noted a possible correlation at the 95\% level,
although this was strongly driven by a single high D/H
sightline.  Moreover, Prochaska et al. note that the literature
limit for a second high D/H sightline ($\gamma^2$ Vel)
does not agree with the overall trend in their data and that there
is a large scatter in Ti/H at low D/H.  
More recently, Linsky et al. (2006) have presented
corroborating evidence for depletion correlations based
on Si and Fe; both elements show significant 
abundance trends with D/H.  Linsky et al. also tentatively
suggest  that a possible correlation between D/H and
the gas temperature derived from H$_2$ may also support
the dust scenario.  Savage et al. (2007) have also supported
the dust depletion model, since their high value (2.2$\times 10^{-5}$) 
of D/H in the warm neutral medium of the Galactic halo is
consistent with expulsion of material from the disk and
associated ablation of dust particles.

\begin{center}
\begin{table*}
\caption{Journal of Observations}
\begin{tabular}{lccccccc}
\hline
Star & Mag (V) &  RA (J2000) & Dec (J2000) & Distance (pc) & Instrument & Run \\
\hline
$\delta$ Ori  &2.23 & 05 32 00.4 & $-$00 17 57 & 281$\pm$65 & HIRES & September 2004  \\
$\iota$ Ori  & 2.77 &05 35 26.0 &$-$05 54 35.6 & 407$\pm$127& HIRES & September 2004 \\
$\epsilon$ Ori  &1.70 &05 36 12.8 &$-$01 12 06.9 & 412$\pm$154 & HIRES & September 2004  \\
$\mu$ Col & 5.17 &05 45 59.9 & $-$32 18 23.2  & 400$^{+100}_{-70}$ & UVES & Archive  \\
HD 41161 & 6.76 &06 05 52.5 &+48 14 57.4 & 1253 & HIRES & April 5 2006  \\
HD 53975 & 6.47 &07 06 36.0 &$-$12 23 38.2 & 1318 & HIRES & April 5 2006   \\
$\zeta$ Pup & 2.25 &08 03 35.1 &$-$40 00 11.3 & 429$\pm$94 & UVES & Period 76   \\
$\gamma^2$ Vel &1.78  &08 09 32.0 &$-$47 20 11.7 & 258$\pm$35 & UVES & Period 76  \\
WD 1034+001 &13.22 & 10 37 04.0 &$-$00 08 20 & 155$^{+58}_{-48}$ & HIRES & June 2006   \\
$\theta$ Car & 2.76 & 10 42 57.4& $-$64 23 40.0& 135$\pm$9 & UVES & Period 76   \\
$\alpha$ Cru & 1.33 & 12 26 35.9 & $-$63 05 56.7& 98$\pm$6 & UVES& Period 76   \\
$\beta$ Cen &0.61  &14 03 49.4 &$-$60 22 22.9 & 161$\pm$16 & UVES & Archive   \\
BD+39$^\circ$3226 & 10.18 &17 46 31.9 &+39 19 09.07 & 290$^{+140}_{-70}$ & HIRES & April 5 2006  \\
HD 191877  & 6.26& 20 11 21.0&+21 52 29.8 & 2200$\pm$550& HIRES & September 2004  \\
HD 195965  &6.98 &20 32 25.6 &+48 12 59.3 & 794$\pm$200 & HIRES & September 2004   \\
HD 206773 & 6.9 &21 42 24.2 &+57 44 10 & 500$^{+200}_{-120}$ & HIRES & June 2006   \\
BD+28$^\circ$4211 &10.51 &21 51 11.0 &+28 51 50.4 & 104$\pm$18 & HIRES & October 2004   \\
HD 209339  &6.69 &22 00 39.3 &+62 29 16.0 & 850 & HIRES & June 2006   \\
Feige 110  &11.83 &23 19 58.4 &$-$05 09 56.2 & 179$^{+265}_{-67}$ & HIRES & September 2004   \\
\hline 
\end{tabular}\label{archive}
\end{table*}
\end{center}

In this paper, we extend previous analyses of refractory
elements by significantly enlarging the sample of Ti measurements
in the local ISM.  Due to its highly refractory nature
(e.g. Savage \& Sembach 1996; Jenkins 2004), Ti should provide
one of the most sensitive measurements of depletion and therefore
correlate most steeply with D/H.  A further advantage of Ti~II
is its ionization potential (IP) of 13.6 eV which means that ionization
corrections will be less than for other commonly detected species
such as Fe~II (IP=16.2 eV) and Si~II (IP=16.3 eV).

\section{Observations and Data Analysis}

We chose targets with measured D/H and whose log N(HI) $>$ 19.7.
This criterion ensures that Ti~II lines will have 
detectable equivalent widths (EWs) 
and that ionization corrections will be a negligible concern.

\subsection{Observations and Data Reduction}

Our final sample of Ti~II lines in the local Galactic disk comprises
data obtained from the following sources:  the pilot survey of
Prochaska et al. (2005) (7 stars), VLT/UVES data obtained from the
archive (2 stars), VLT/UVES data obtained through a Period 76
ESO allocation (4 stars) and new HIRES observations obtained in
2006 (6 stars).
For completeness, we list all targets from our Ti~II survey (which
include those already published in Prochaska et al. 2005) in Table 
\ref{archive}, but only describe the data acquisition of previously
unpublished data in the following sections.  The distances to the
stars are taken from the compilation of Linsky et al. (2006),
supplemented by values from Pan et al. (2004), 
Oliveira \& Hebrard (2006) and Hebrard et al. (2005a).

\subsubsection{HIRES Data}

The new HIRES data were obtained, primarily during twilight, 
on the nights of 06 April 2006, 02 June 2006, 03 June 2006, 
and 25 December 2006.  All of the data were acquired with the
0.4$''$ wide, E3 decker which affords a FWHM resolution of 
$\approx 3$\kms.  We reduced these data using the HIRedux pipeline
(v. 2.1; Bernstein et al., in preparation) developed using the IDL software 
package and distributed within the XIDL package
(http://www.ucolick.org/$\sim$xavier/IDL/index.html).
In brief, the pipeline bias subtracts and flatfields
each science and calibration image.  It traces the order curvature
across each chip (limited to the blue CCD for these observations)
and derives a 2D wavelength solution from the ThAr images. 

For these very bright targets, the biggest background is scattered
light.  This component is fit with a 2D surface by measuring
the signal between echelle orders.  The fit is interpolated across the orders
and subtracted from the image.  The
RMS error in this signal is typically less than 5~electrons.  
We then boxcar extracted the stars (the S/N is too large for optimal
extraction to be advantageous) to derive a 1D spectrum and 
co-added the individual exposures, weighting by the median S/N.

\subsubsection{Archival UVES Data}

Data for $\mu$ Col and $\beta$ Cen were downloaded from the ESO
archive.  A considerable amount of data were available but we
selected a subset whose settings offered the highest resolution
and maximum exposure times at the wavelengths of interest.  For
$\beta$ Cen we used only 1x1 binned data with 4 central wavelengths:
346 nm (slit width = 0.5 arcsec), 390 nm  (slit width = 0.44 arcsec),
437 nm  (slit width = 0.5 arcsec) and 564 nm  (slit width = 0.3 arcsec)
yielding a FWHM resolution ranging from $R \sim 70,000$ to 90,000
depending on the slit and the CCD. For $\mu$ Col,
the data were more limited and often had sub-optimal instrument set-ups
(such as very wide slits).  We therefore used only 1 central wavelength 
setting, 390 nm, with 2x2 binning and a 1 arcsec slit
yielding a FWHM resolution $R \sim 45,000$.  Although a set of
MIDAS scripts exists to form a UVES pipeline, we found that the pipeline
reduction had considerable trouble dealing with very high S/N data
(a well-known problem).  For consistency, we therefore
decided to apply the HIRedux IDL reduction pipeline to the UVES data.
This process worked well in general, although the final extracted
spectra had considerable structure in the continuum (perhaps owing to 
poor flat-fielding), particularly in the
blue orders.  We attempted to fit out this structure during the continuum
fitting process.  This process was judged to be quite successful,
based on the consistency of EWs of a given interstellar species
between different orders, settings and transitions (see next section).

\subsubsection{UVES Period 76 Data}

\begin{figure}
\centerline{\rotatebox{0}{\resizebox{9cm}{!}
{\includegraphics{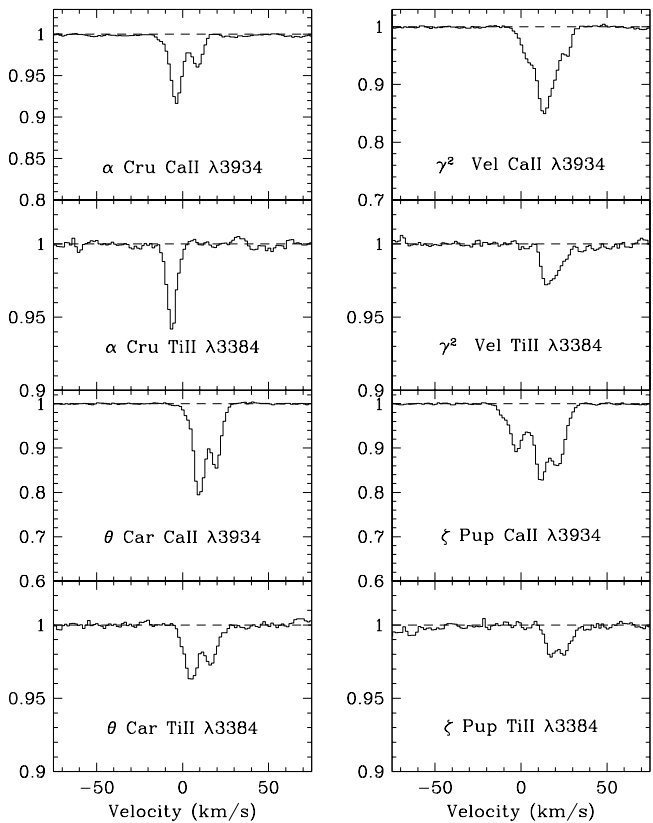}}}}
\caption{\label{lines1} Normalized sections of spectra around the
transitions of Ti~II $\lambda$ 3384 and Ca~II $\lambda$ 3934 for
4 of our program stars (see also Figure \ref{lines2} and \ref{lines3}).  
Note the different y-axes from panel to panel. }
\end{figure}

\begin{figure}
\centerline{\rotatebox{0}{\resizebox{9cm}{!}
{\includegraphics{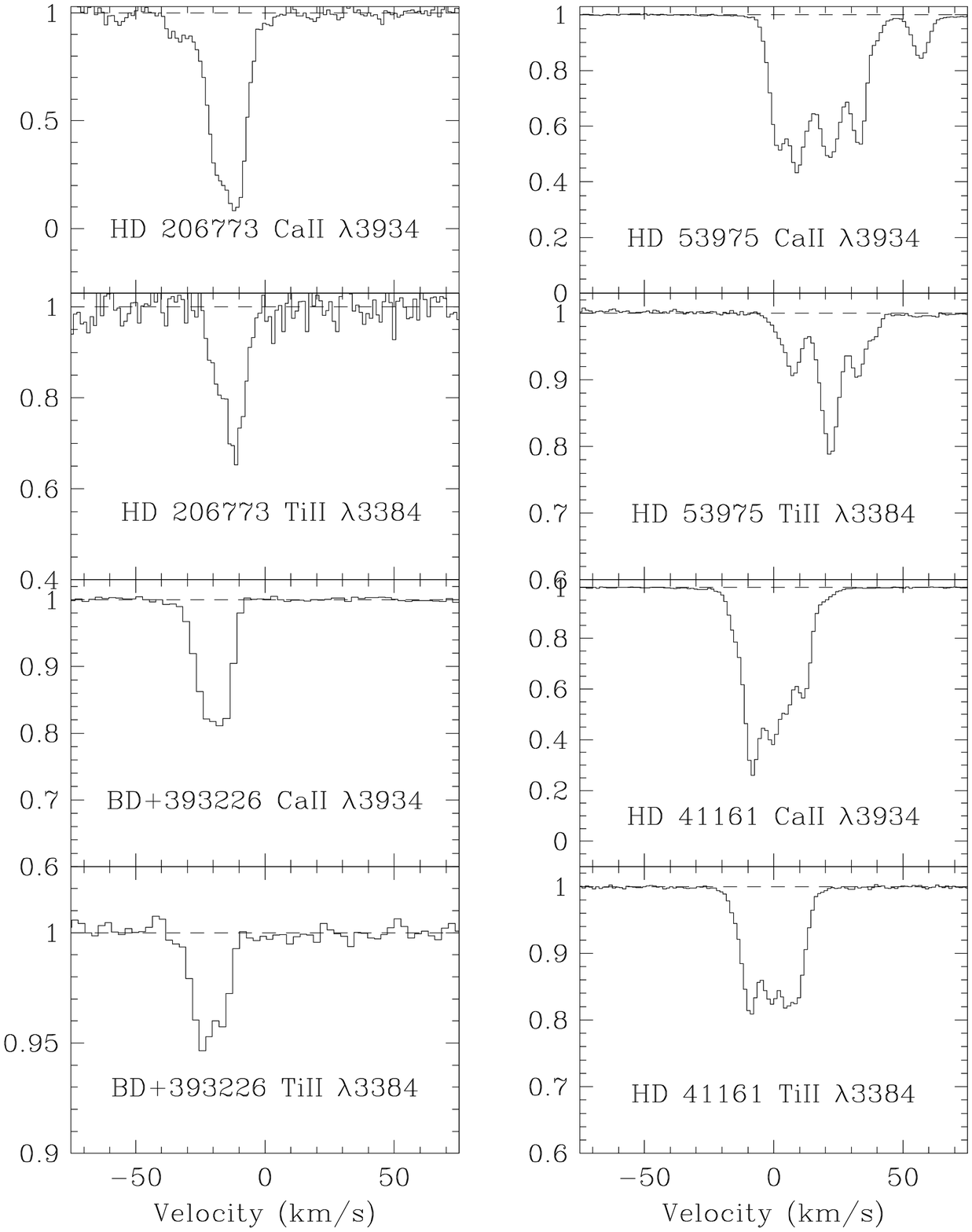}}}}
\caption{\label{lines2} Normalized sections of spectra around the
transitions of Ti~II $\lambda$ 3384 and Ca~II $\lambda$ 3934 for
4 of our program stars (see also Figures \ref{lines1} and \ref{lines3}).  Note the
different y-axes from panel to panel. }
\end{figure}

\begin{figure}
\centerline{\rotatebox{0}{\resizebox{9cm}{!}
{\includegraphics{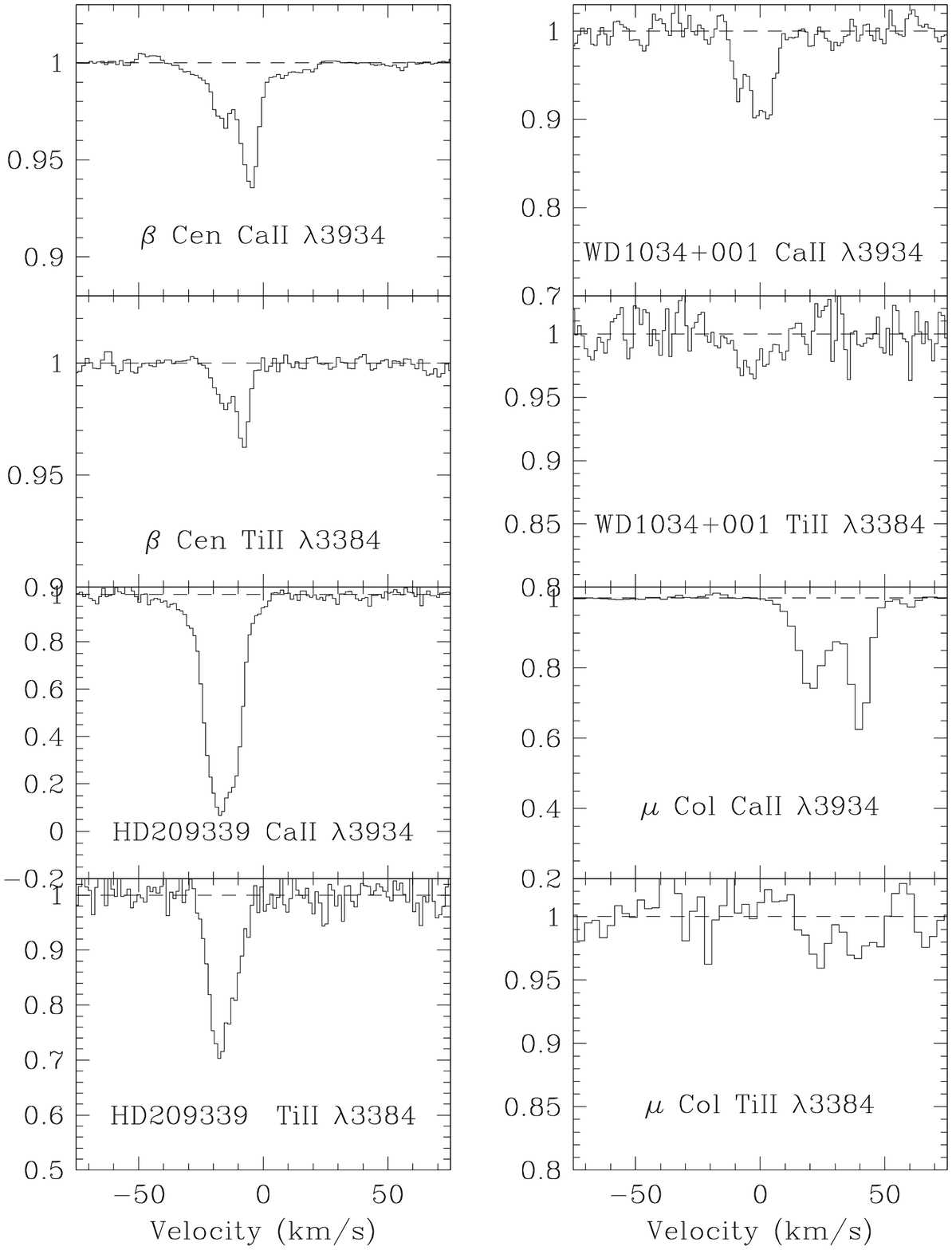}}}}
\caption{\label{lines3} Normalized sections of spectra around the
transitions of Ti~II $\lambda$ 3384 and Ca~II $\lambda$ 3934 for
4 of our program stars (see also Figure \ref{lines1} and \ref{lines2}).  Note the
different y-axes from panel to panel. }
\end{figure}

For 4 southern stars not available in the archive, we received an
ESO allocation in Period 76 (October 2005 -- March 2006).  In all
cases we used the same combination of settings: 1x1 binning, 0.5 arcsec
slit and 3 central wavelengths, 346 nm, 390 nm and 580 nm
yielding a typical resolution $R \sim 80,000$.  As for the
archival UVES data, we used the adapted version of the HIRES
IDL pipeline and fitted out the structure in the blue orders
along with the underlying continuum.

\section{RESULTS}

\begin{table*}
\caption{Column Density Data}
\begin{tabular}{lcccccccl}
\hline
Star & N(H) & N(D) & N(OI) & N(TiII) & N(CaII)  & N(FeII) & N(SiII) & Refs \\
\hline
$\delta$ Ori & 20.19$\pm$0.03 & 15.06$\pm$0.12  & 16.67$\pm$0.05 & 11.15$\pm$0.04 & ...  &  14.08$\pm0.03$  &  ... & 1, 2, 3, 4 \\
$\iota$ Ori & 20.18$^{+0.06}_{-0.07}$ & 15.30$^{+0.04}_{-0.05}$ & 16.76$\pm$0.09 & 11.31$\pm$0.03 &  ... &  14.20$^{+0.20}_{-0.15}$ & 15.16$^{+0.02}_{-0.03}$ & 1, 3, 4, 5\\
$\epsilon$ Ori & 20.45$\pm$0.08 & 15.25$\pm$0.05 & 16.98$\pm$0.05 & 11.40$\pm$0.03 & ...  &  14.20$\pm0.10$ & 15.02$^{+0.08}_{-0.12}$ & 1, 3, 4, 5\\
$\mu$ Col & 19.86$\pm$0.02 & 14.7$^{+0.3}_{-0.1}$ & ... & $<$11.4 & 12.13$\pm$0.02  &14.13$\pm0.02$ & 15.10$\pm0.02$&4, 6\\
HD 41161 & 21.08$\pm$0.09 & 16.41$\pm0.05$ & 18.04$\pm0.05$ & 12.28$\pm$0.04 & 12.54$\pm$0.01 &   14.98$\pm0.06$ & ... & 7 \\
HD 53975 & 21.14$\pm0.06$ & 16.15$\pm0.07$ & 17.87$\pm0.08$ & 12.13$\pm$0.04 & 12.56$\pm$0.01 &   14.84$\pm0.05$ & ... & 7 \\
$\zeta$ Pup & 19.96$\pm$0.03 & 15.11$\pm$0.06 & ... & 11.07$\pm$0.04 &11.83$\pm$0.02  &  14.13$^{+0.17}_{-0.07}$ & 15.07$^{+0.07}_{-0.02}$ & 4, 8 \\
$\gamma^2$ Vel & 19.71$\pm$0.03 & 15.05$\pm$0.03 & ... & 11.10$\pm$0.02 & 11.61$\pm$0.02 &   13.95$\pm0.11$ & 15.03$\pm0.04$ & 4, 8 \\
WD 1034+001 & 20.07$\pm$0.07 & 15.40$\pm$0.07  & 16.60$\pm$0.10 & 11.1$\pm$0.2  & 11.35$\pm0.15$  &   14.10$\pm0.10$ & ...& 4, 9 \\
$\theta$ Car & 20.28$\pm$0.10 & 14.98$^{+0.18}_{-0.21}$ & 16.42$\pm$0.20 & 11.31$\pm$0.01& 11.65$\pm$0.01 &    14.19$\pm0.03$ & 14.38$^{+0.25}_{-0.08}$ & 4, 10 \\
$\alpha$ Cru & 19.85$\pm$0.10 & 14.95$\pm$0.05 & ...& 11.12$\pm$0.02 & 11.23$\pm$0.05 &   14.00$\pm$0.10& ... & 4, 6 \\
$\beta$ Cen & 19.63$\pm$0.10 & 14.7$\pm$0.2 & ... & 11.05$\pm$0.04 & 11.12$\pm$0.04 &   13.92$^{+0.05}_{-0.04}$ & 14.59$\pm0.01$ & 4, 6, 11\\
BD+39$^\circ$3226 & 20.08$\pm$0.09 & 15.15$\pm$0.05 & 16.40$\pm$0.10 & 11.49$\pm$0.08  & 11.67$\pm$0.01  &   14.15$\pm0.07$ & 14.80$\pm0.20$ & 4, 9\\
HD 191877 & 21.10$\pm$0.07 & 15.95$^{+0.11}_{-0.06}$ & 17.54$^{+0.20}_{-0.12}$ & 12.24$\pm$0.02 &  ...  & 14.95$\pm0.02$ & ...& 3, 4, 12, 13\\
HD 195965 & 21.01$\pm$ 0.05 & 15.89$\pm$0.07 & 17.77$^{+0.04}_{-0.06}$ & 12.02$\pm$0.02 & ...   & 14.81$\pm0.01$ & ... & 3, 4, 12, 13\\
HD 206773$^\dagger$ & 21.10$\pm$0.05 &   17.0$\pm$0.5  & 17.85$\pm$0.05 & 12.16$\pm$0.02 &12.65$\pm$0.02   & ... & ... & 14 \\
BD +28$^\circ$4211 & 19.85$\pm$0.02 & 14.99$\pm$0.03 & 16.22$\pm$0.10 & 11.08$\pm$0.08 &... &  14.10$\pm0.10$ & ...& 3, 4, 15\\
HD 209339$^{\dagger\dagger}$& $\sim$21.5 &  ...   & ... &  12.14$\pm$0.03& 12.68$\pm$0.03  &   ... & ... &  16\\
Feige 110 & 20.14$\pm$0.07 & 15.47$\pm$0.06 & 17.06$\pm$0.15 & 11.59$\pm$0.03 & ...  &   ... & ... & 3, 17, 18 \\
\hline 
\end{tabular}\label{results}
\\ Upper limits are 3$\sigma$. $^\dagger$ Included here for completeness, but not used in analysis due to
saturation of DI lines (Hebrard et al. 2005a).$^{\dagger\dagger}$  Included here for
completeness, FUSE deuterium data still under analysis (C. Oliveira, private communication).  Refs:
1 - Meyer et al. (1998); 2 - Jenkins et al. (1999); 3 - Prochaska et al. (2005); 4 - Linsky et al. (2006);
5 - Laurent et al. (1979); 6 - York \& Rogerson (1976); 7 - Oliveira \& Hebrard (2006);
8 - Sonneborn et al (2000); 9 - Oliveira et al. (2006); 10 - Allen et al (1992); 11 - van Steenberg \& Shull (1988);
12 - Hoopes et al. (2003); 13 - Liszt (2006); 14 - Hebrard et al. (2005a); 15 - Sonneborn et al. (2002); 16 - C. Oliveira, private communication;
17 - Friedman et al. (2002); 18 - Hebrard et al. (2005b)
\end{table*}

\subsection{Analysis}\label{analysis_sec}

To determine gas phase column densities we use the 
VPFIT\footnote{http://www.ast.cam.ac.uk/\~{}rfc/vpfit.html }
software to decompose the absorption into individual components.
The total column density is then calculated from the sum of these
components. Profiles of Ti~II $\lambda$ 3384, plus Ca~II $\lambda$3934
are shown in Figures \ref{lines1}, \ref{lines2} and \ref{lines3}.  
In many lines of sight, Ti~II $\lambda$
3242 is also detected, in which case we simultaneously fit the two transitions.
The very small EWs (typically 10-20m\AA) of the Ti~II
lines coupled with the $b$-values (which indicate that the lines
are resolved) show that it is unlikely that saturation affects 
these transitions.
This is confirmed by an optical depth analysis of the two Ti~II
lines which yields consistent results for both.
In the UVES data and a small subset of the HIRES data, 
Ca~II $\lambda$ 3969 is also covered so that, when
detected, it is simultaneously fit with Ca~II $\lambda$ 3934.
In the case of WD1034+001, the
Ti~II and Ca~II lines are barely detected, so the fit is very uncertain.  
In this case, we therefore additionally estimate the column density from the
apparent optical depth of the absorption lines.  For $\mu$ Col,
the Ti~II $\lambda$3384 line is detected at $< 3 \sigma$ significance,
so we quote a limit.  We determine the errors
on the lines in two different ways.  First, we consider the formal
error on the total column density returned by VPFIT\footnote{The
errors on individual component column densities are also returned,
but these are considerably more unreliable.}.  Next, we separately
fit the different transitions of the same species, as well as fitting
lines of a given species that are covered by multiple orders,
or multiple settings.  This is particularly useful for the assessment
of systematic errors such as the continuum fit, a particular concern
for the blue UVES data.  However, we found that the separate fits
exhibited excellent agreement, usually within 0.02 dex.  Due to
the very high S/N of the spectra, the range of column densities
determined from individual line fits usually dominated the
formal error from VPFIT, so that the former was adopted as the
final quoted uncertainty in Table \ref{results}.
We exclude two of the sightlines from our analysis of depletion patterns.
First, HD 206773 has poorly determined D/H due to severe saturation
of the D~I lines (Hebrard et al. 2005a).  Second, 
although HD 209339 has FUSE data
which will eventually permit an analysis of the D/H in this
sightline, the data have not yet been analysed.  We tabulate
the column densities of Ti~II and Ca~II for completeness, but
do not consider these sightlines further in our analysis.

For metal line species, we quote in Table \ref{results} the
column densities of the single atomic or ionic species that is
measured in our (or literature) data.  However, Liszt (2006)
has shown that the molecular contribution to D and H can be
important and that ignoring its contribution can lead to an
over-estimate of the D/H ratio (see also Lacour et al. 2005).  
We therefore use total column densities for hydrogen
and deuterium, accounting for both molecular and atomic gas, 
i.e. N(H)=N(HI)+2N(H2)+N(HD)
and N(D) = N(DI)+N(HD).  For most sightlines, the molecular
contribution is orders of magnitude less than the atomic column
density such that N(H) $\sim$ N(H~I) and N(D) $\sim$ N(D~I).
However, a significant effect is present for HD~41161, HD~53975,
HD~191877, HD~195965, HD~206773 and HD~209339 (although these
latter two are not used in our main analysis).
The largest effect is seen towards HD~41161 where D~I/H~I = 25.1 parts
per million (ppm) and
D/H = 21.4 ppm (Oliveira \& Hebrard 2006), a difference of $\sim$ 15\%.

\subsection{Correlations with D/H}

\begin{center}
\begin{table*}
\caption{Abundances}
\begin{tabular}{lccccc}
\hline
Star & (D/H)$_{\rm ppm}$ & [Ti/H] & [Si/H] & [Fe/H] & log \avgnh \\
\hline
$\delta$ Ori  & 7.4$^{+2.4}_{-1.8}$ & $-1.96\pm$0.05 &  ... &  $-1.58\pm0.04$ &  $-0.75\pm0.04$ \\
$\iota$ Ori  & 13.2$^{+2.6}_{-2.2}$ & $-1.76\pm$0.07 &  $-0.53\pm0.06$ &  $-1.42\pm0.18$ & $-0.95\pm0.04$ \\
$\epsilon$ Ori  & 6.3$^{+1.5}_{-1.2}$ & $-1.97\pm0.09$ & $-0.97\pm0.13$ & $-1.72\pm0.13$ & $-0.65\pm0.08$ \\
$\mu$ Col &  6.9$^{+4.1}_{-2.6}$ &  $<-1.38$ &  $-0.30\pm0.03$ & $-1.20\pm0.03$ & $-1.23\pm0.01$ \\
HD 41161 & 21.4$^{+5.7}_{-4.5}$ & $-1.72\pm0.10$ &  ... & $-1.57\pm$0.11 &  $-0.51\pm0.08$ \\
HD 53975 &  10.2$^{+2.4}_{-2.0}$ & $-1.93\pm0.07$ &  ... & $-1.77\pm0.08$ &  $-0.47\pm0.09$ \\
$\zeta$ Pup & 14.1$^{+2.4}_{-2.0}$ & $-1.81\pm0.05$ &  $-0.43\pm0.06$ & $-1.30\pm0.10$ & $-1.16\pm0.02$ \\
$\gamma^2$ Vel & 21.9$^{+2.2}_{-2.0}$ & $-1.53\pm0.05$ & $-0.22\pm0.05$ & $-1.23\pm0.11$ & $-1.19\pm0.01$ \\
WD 1034+001 & 21.4$^{+5.5}_{-4.4}$ & $-1.89\pm0.21$ &  ... &  $-1.44\pm0.12$ & $-0.61\pm0.08$ \\
$\theta$ Car & 5.0$^{+3.4}_{-2.0}$ & $-1.89\pm0.10$ & $-1.44\pm0.22$ & $-1.56\pm0.10$ & $-0.34\pm0.03$ \\
$\alpha$ Cru & 12.6$^{+3.7}_{-2.9}$ & $-1.65\pm0.10$ & ... & $-1.32\pm0.14$ & $-0.63\pm0.01$ \\
$\beta$ Cen & 11.7$^{+7.9}_{-4.7}$ & $-1.50\pm0.11$ & $-0.58\pm0.10$ & $-1.18\pm0.11$ & $-1.07\pm0.01$ \\
BD+39$^\circ$3226 & 11.7$^{+3.1}_{-2.5}$ & $-1.51\pm0.12$ & $-0.82\pm0.22$ & $-1.40\pm0.11$ & $-0.87\pm0.05$ \\
HD 191877  & 7.1$^{+2.3}_{-1.7}$ & $-1.78\pm0.07$ &  ... & $-1.62\pm0.07$ & $-0.73\pm0.05$ \\
HD 195965   & 7.6$^{+1.7}_{-1.4}$ & $-1.91\pm0.05$ & ... & $-1.67\pm0.05$ & $-0.38\pm0.11$ \\
BD+28$^\circ$4211 & 13.8$^{+1.2}_{-1.1}$ & $-1.69\pm0.08$ &  ... &  $-1.22\pm0.10$ & $-0.66\pm0.04$ \\
Feige 110   & 21.4$^{+5.1}_{-4.1}$ & $-1.47\pm0.08$ &  ... & ... &    $-0.60\pm0.35$ \\
\hline 
\end{tabular}\label{abund}
\end{table*}
\end{center}

\begin{figure}
\centerline{\rotatebox{0}{\resizebox{8cm}{!}
{\includegraphics{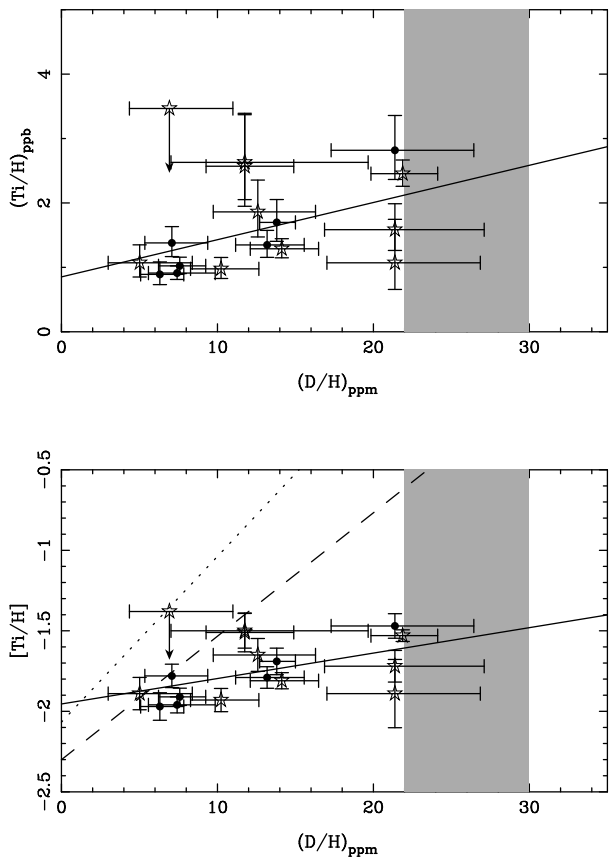}}}}
\caption{\label{tih} Ti abundance as a function of D/H.  Results from
this work are show by stars; filled circles are data points taken from 
Prochaska et al. (2005).  The solid line shows the least squares
best fit to all 16 detections.  The shaded region shows the constraints
on the primordial value of D/H determined from quasar absorption line
systems.  In the top panel, we show linear-linear
axes for Ti/H (in parts per billion, ppb)
and D/H (in parts per million, ppm), 
the best fit line $y = mx +c$ has $m$=0.06$\pm$0.03
and $c$=0.85$\pm$0.32.  In the lower panel, we us a logarithmic y-axis
to facilitate comparison with other elements, the best fit line 
$y = mx +c$ has $m$=0.016$\pm$0.006
and $c = -1.95\pm0.08$.  The fits to Fe and
Si abundances from Linsky et al. (2006) are shown with dashed and
dotted lines respectively.  }
\end{figure}

In Figure \ref{tih} we present the first main result of this analysis.
The abundance of Ti, on both linear (in units of parts per billion, ppb)
and logarithmic (relative to solar) scales is
plotted versus D/H in units of ppm. For logarithmic
abundances, we use the standard notation 
[X/H] = log (X/H) $-$ log (X/H)$_{\odot}$, which is analogous
to the notation D(X) used by Linsky et al. (2006).  For solar abundances,
we use the compilation of Lodders (2003), 
except for oxygen, for which we adopt the value from Holweger (2001)
where a full non-LTE treatment was adopted.  
These values are log [N(X)$_{\odot}$/N(H)$_{\odot}$] = 
$-4.46, -4.53, -7.08, -3.26$ for Si, Fe,
Ti and O respectively. For convenience, in Table \ref{abund} we show the 
relative abundances used for Figures \ref{tih} and \ref{nh_fig}. 
We include data
from Prochaska et al. (2005) and perform a least-squares fit
of the form $y = mx + c$ to the final Ti sample which consists of 
16 detections.  
The best fit straight line for [Ti/H] has values $m=0.016\pm0.007$ and 
$c=-1.95\pm0.09$ where errors were determined by randomly re-generating
the [Ti/H] vs. D/H distribution 1000 times based on the quoted 1
$\sigma$ errors.
Spearman's rank correlation statistic is 0.55, yielding
a rejection of the null hypothesis (that there is no correlation)
at the 97\% level.  In addition to the enlargement of the sample size,
one of the
important contributions of this work is the inclusion
of 3 new high D/H data points, as well as the lowest D/H in the
current sample.  The correlation of Prochaska et al.
(2005) was pivotal upon a single high datum (Feige 110); removal of this one
point reduces the significance of the correlation from 95\% to
86\%.  Moreover, for one other high D/H sightline in the literature,
Prochaska et al. 
(2005) pointed out that the upper limit for $\gamma^2$ Vel (D/H=21.9 ppm)
of [Ti/H]$\le-1.82$ reported by Welsh et al. (1997) was potentially
inconsistent with the depletion trend.  With a larger sample, and more
high D/H points, we can now make a more robust assessment of the D/H 
correlation.  First, we find that with higher S/N data,
the measured value of [Ti/H]=$-1.53$ for  $\gamma^2$ Vel 
is in actually excellent agreement with
the high D/H point (Feige 110) presented in Prochaska et al.
Moreover, the highest D/H sightline (HD41161) in our sample is also in
good agreement with the general trend of increasing Ti/H with
D/H.  However, the third high D/H sightline, WD1034+001 
exhibits a low Ti/H for its D/H, although this point has the
largest error bars on its measured N(Ti~II) column density
due to the low significance of the detection (see Figure \ref{lines3}
and Tables \ref{results} and \ref{abund}).

Given the discrepancy noted above between the literature upper
limit and our measured value for [Ti/H] towards $\gamma^2$ Vel,
it is interesting to compare other sightlines for which
literature values also exist. Welsh et al. (1997) report a limit on the
column density of Ti~II towards $\theta$ Car that is half of
our measured value and the discrepancy towards $\zeta$ Pup is
even higher: 0.45 dex, almost a factor of three.  
However, the Ca~II abundances in the 
sightlines common to this work and Welsh et al. (1997) are
in good agreement within the errors, with differences $\le$ 0.08 dex.  
Hunter et al. (2006) have also previously
noted consistency in N(Ca~II) and discrepancies in N(Ti~II) 
between high S/N UVES data and previous
works, whose S/N ratios are typically $<$ 50.  Hunter et al. attributed
this to the weakness of the Ti~II lines, relative to Ca~II.
Confirmation of this effect here highlights the importance
of very high S/N ratios and spectral resolution
to make these measurements accurately.

\section{Discussion}

\subsection{Dust Depletion}

Previous authors have interpreted the correlation between 
the gas-phase abundance of refractory elements and (D/H) 
as evidence that differential depletion leads to low values
for each (Prochaska et al.\ 2005; Linsky et al.\ 2006).
Although the trend traced by our enlarged sample of Ti/H values
supports this interpretation qualitatively, a closer examination
of the results leads to a less clear cut picture.
Focus first on the abundance trends from different elements.
In Figure \ref{tih} we overlay the fits for depletion of
Fe and Si from Linsky et al. (2006) with our own Ti results.  
If dust depletion is the
cause of the correlation of abundance with D/H then we expect
a steeper gradient for more refractory elements.  In the case
of Ti, Fe and Si, the former is the most refractory and the latter
the least sensitive to depletion (e.g. Savage \& Sembach 1996).
The relative gradients in Figure \ref{tih} are therefore inconsistent
with our simple predictions from a `standard' depletion scenario, 
since we would expect Ti to be
considerably steeper than Fe or Si.  Indeed, the gradient seen in
Ti is only marginally steeper than that found for [O/H] which
is only very mildly refractory.  Therefore, although the
existence of trends between refractory elements and D/H
support a picture of dust depletion, quantitatively the gradients
are inconsistent with this scenario.  
Steigman, Romano \& Tosi (2007) have also questioned the dust depletion
scenario due to their failure to find a trend between
line of sight reddening and D/H, as one might have expected were
dust the cause of lower D/H values.  

In section \ref{analysis_sec} we argued that saturation of the Ti~II
lines is unlikely to be a serious concern.  However, some of the Fe~II and
Si~II transitions have considerably higher optical depths (e.g.,
Redfield \& Linsky 2002).  There is an anti-correlation between
[Fe/H] and N(Fe~II) in the compilation of Linsky et al. (2006).
If saturation systematically under-estimates N(Fe~II) for the low 
[Fe/H] data points,  the true slope of the correlation with D/H 
will be flatter than observed.  However,
this effect is unlikely to be large enough to yield a true
gradient of [Fe/H] versus D/H that is flatter than the [Ti/H]
versus D/H correlation.  Moreover, there is a mild correlation
between [Si/H] and N(Si~II), so that any correction for saturation
would make the dotted line in Figure \ref{tih} steeper, thus accentuating
the difference with the [Ti/H] correlation.  
It is also possible that the D lines may suffer from saturation,
and this may be particularly relevent for the high N(HI) lines
of sight studied here.  However, it is difficult to apply horizontal
shifts to the D/H values in our sample that would make the correlation
with [Ti/H] steeper.
It therefore seems unlikely that saturation effects can explain
the relative slopes in Figure \ref{tih}.

We can further investigate depletion trends by considering correlations
with mean hydrogen densities. It is well known that
depletion in the Galactic ISM correlates with \avgnh, the
mean volume density of hydrogen (e.g. Savage \& Bohlin 1979;
Phillips, Gondhalekar \& Pettini 1982;
Spitzer 1985; Gondhalekar 1985;  Jenkins et al. 2004) where

\begin{equation}
<n(H)> = \frac{N(HI) + 2N(H_2)}{\rm distance}
\end{equation}

\begin{figure*}
\centerline{\rotatebox{270}{\resizebox{10cm}{!}
{\includegraphics{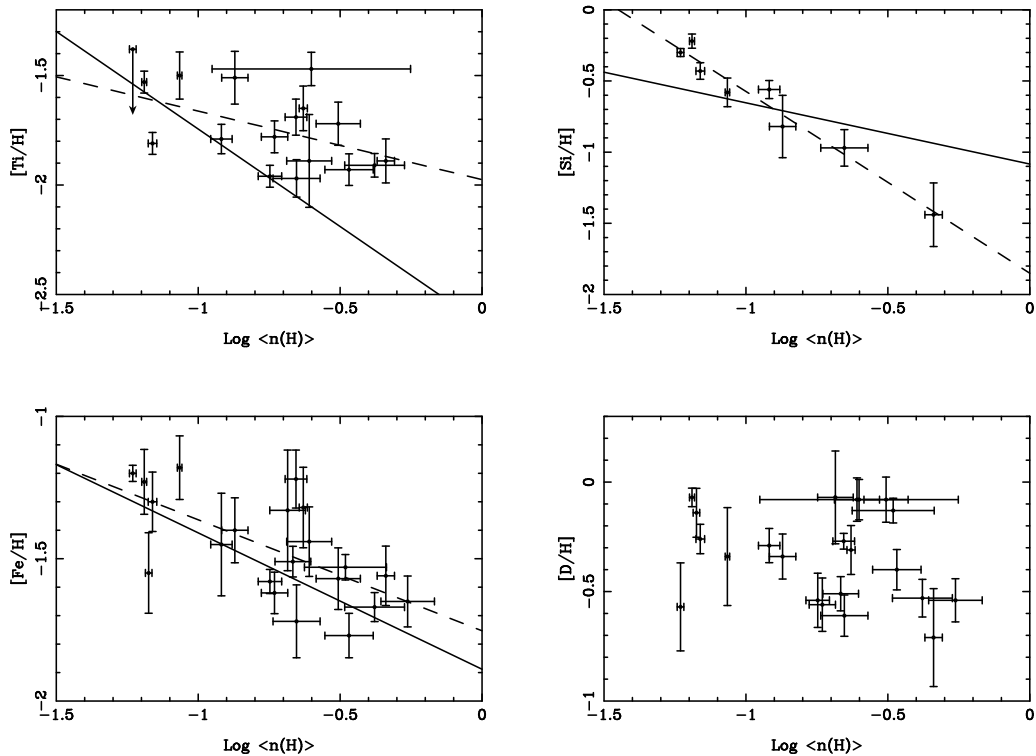}}}}
\caption{\label{nh_fig} Abundances versus total hydrogen volume density, n(H)
for a sample of local Galactic sightlines from Linsky et al. (2006) and this
paper.  Whereas Ti, Si and Fe are plotted relative to solar, the 
deuterium abundances 
are plotted relative to the primordial value, which we take to be 26 ppm.
The solid lines show a fit to [X/H] vs. \avgnh\ determined from
144 stellar sightlines (Jenkins 2004). The dashed lines are the least
squares fits to the data presented and have equations:
[Ti/H]= $-0.31$\avgnh\ $-1.98$, [Si/H]= $-1.28$\avgnh\ $-1.86$, 
[Fe/H]= $-0.39$\avgnh\ $-1.75$. }
\end{figure*}

We calculate \avgnh\ for the 17 sight lines in our Ti~II sample
and supplement them with five more sight lines in Linsky et
al. (2006)  which have H$_2$ measurements (Lan~23, TD1~32709, PG0038+199,
LSS~1274, and HD~90087) see Table \ref{abund} and Figure \ref{nh_fig}.
Jenkins (2004) parametrized ISM abundances in terms of a depletion
factor, $F_{\star}$, which in turn correlates with \avgnh.  We performed a
fit to the correlation between $F_{\star}$ and \avgnh\ and using the
parameters in Jenkins (2004), determined the corresponding relationship
between \avgnh\ and abundance for the 144 Galactic stars in his sample.
For Ti, Si and Fe we determine:

\begin{equation}
\rm[Ti/H]= -2.226 (0.4<n(H)> + 0.8) - 0.844 - 0.01
\end{equation}

\begin{equation}
\rm[Si/H] = -1.076 (0.4<n(H)>+0.8) - 0.223
\end{equation}

\begin{equation}
\rm[Fe/H] = -1.198 (0.4<n(H)> +0.8) - 0.95 + 0.02
\end{equation}

The final term in the equations for [Ti/H] and [Fe/H] account for
the slightly different solar abundance scale adopted by Jenkins (2004).
This relationship between \avgnh\ and ISM abundances is over-plotted
with our data in Figure \ref{nh_fig}.
The correlation between [Ti/H] and depletion factor
in Jenkins (2004) is very steep (the steepest of the 15 elements
he studies), and very tight (reduced $\chi^2$=0.95). The best
least squares fit to our Ti data is shown by the dashed line
in the top left panel of Figure \ref{nh_fig}.  The slope is
clearly much flatter than that reported Jenkins (2004) and flatter
than both the least squares fits for Fe and Si.  Moreover,
the correlation between [Ti/H] and \avgnh\ is not statistically
significant: the Spearman rank correlation coefficient rules
out the null hypothesis at $< 2 \sigma$ (83\%) significance.   
It is also interesting to note that the abundance of Si, which is the least
depleted of the three elements discussed here, has the steepest gradient
with \avgnh.  This reflects the same, unexpected result presented in
Figure \ref{tih}, i.e. that the depletion fractions seem to be
inverted relative to our expectations.  This may indicate
that the normal circumstances which govern the correlation between
depletion and density in the ISM at large is not at work in these
local (mostly $d < 500$ pc) sightlines.    
However, we should also stress that we have many
fewer sightlines than the Jenkins sample, so that small
number statistics could contribute to the absence of a tight correlation.
Also noteworthy is the lack of correlation, and very large
scatter at any given density, of [D/H]\footnote{Here, [D/H] represents
the abundance relative to the primordial value, which we take to be
26 ppm.}. We have only plotted in Figure  \ref{nh_fig} those stars
in our Ti~II sample.  The same plot, but with a much larger sample
extending to lower values of \avgnh,
has been recently presented by Oliveira et al. (2006) who
note that the scatter in D/H increases dramatically when \avgnh\ $>$ 0.1,
the regime in which most of our points lie.
If lower D/H values do result from differential depletion,
it appears from our Figure  \ref{nh_fig} that this depletion does not 
correlate with \avgnh.   Oliveira et al. (2006) have attempted to
fit an abundance/depletion model to their larger dataset and claim
an anti-correlation between D/H and \avgnh.  However, the reduced
$\chi^2$ for this fit is large (5.8), the scatter in D/H for a given
\avgnh\ is $\sim$ 1 dex (similar to what is seen in our data)
and the anti-correlation driven by just 2 data points.  Indeed, some of the 
highest D/H values occur at the highest \avgnh.  We therefore conclude
that there is no convincing evidence for an anti-correlation between
\avgnh\ and D/H as one might expect from dust depletion.

\begin{figure}
\centerline{\rotatebox{270}{\resizebox{6cm}{!}
{\includegraphics{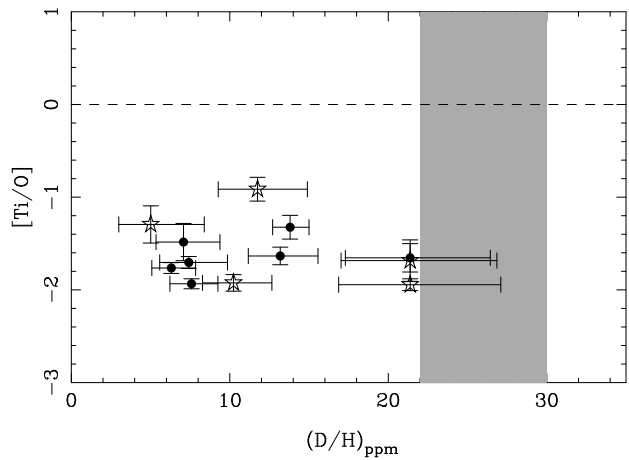}}}}
\caption{\label{tio} Relative Ti/O abundance as a function of D/H.  
Results from
this work are show by stars; filled circles are data points taken from 
Prochaska et al. (2005).   The shaded region shows the constraints
on the primordial value of D/H determined from quasar absorption line
systems.  For two elements with a similar nucleosynthetic origin,
but different refractory properties, we expect a correlation with D/H
if depletion onto dust is the cause of scatter in deuterium abundances.}
\end{figure}

Although the acceptance of dust depletion as the cause of the
local variations of deuterium is gaining momentum, our results
indicate that this interpretation is not clear cut.  The same
conclusion has also been recently reached by Steigman et al. (2007)
who consider the evidence for dust depletion as `ambiguous'.
However, the above discussion assumes a very simple approach to
the behaviour of dust in the local ISM.  There are several caveats
that should therefore be applied, such as the many flavours of
dust that exist with different chemical compositions and the
fact that dust production and destruction is likely to be a dynamic
process.  Moreover, current models of deuterium depletion involve
D incorporation into deuterated PAHs (Draine 2004, 2006), 
different again from the grains which adsorb most
refractory elements.
Combined with the sightline averaged quantities that we
are necessarily dealing with, some scatter in abundance relations
is perhaps not surprising and we return to this issue in section 
\ref{scatter_sec}.  Shocks may also disrupt `normal' depletion
patterns.  For example, weak shocks may be sufficient to release
D from the dust phase (since deuterated PAHs are relatively fragile), 
whereas much stronger shocks may be required
to ablate Ti grain material.  These issues highlight the complexity
of the ISM and abundance ratios and the caution that should be 
exercised in the interpretation that they are driven by dust.

\subsection{Infall}

An alternative scenario to the depletion of deuterium onto grains
is the infall of unenriched gas from the halo which dilutes the
abundance of chemical elements.  Knauth, Meyer \& Lauroesch (2006)
have recently suggested local infall to explain the difference
in N/O abundances at $d > 500$ pc, compared with closer sightlines. 
Romano et al. (2006) have also combined infall with astration
to result in the relatively modest destruction ($f$ = (D/H) $_{\rm prim}$/
(D/H)$_{\rm LISM} \la 1.8$) of deuterium from its primordial value.
In most \textit{simple} infall scenarios, the infalling gas has a D/H abundance
that is close to the primordial value (i.e. higher than typical
ISM values), whilst the abundances of other elements are lower
than in the ambient disk ISM.  This would result in an anti-correlation
between heavy element abundance and D/H (e.g. Steigman et al. 2007), 
rather than the positive correlations depicted in Figure \ref{tih}.  
Infall of low metallicity material could only
explain the positive correlations 
seen in Figure \ref{tih} if the infalling
low metallicity gas is also poor in deuterium.  Although unlikely,
we briefly explore this possibility here.

One might 
discriminate between the two scenarios of dust depletion versus 
various infall scenarios
by examining relative abundances of different heavy elements.
If infall of unenriched (or low metallicity, but with a solar abundance
pattern) gas is responsible for the spatial
variations in D/H, the relative abundances of two elements with
similar nucleosynthetic origins should not correlate with D/H.
Conversely, if the same two elements have different refractory
properties, they should vary with D/H if depletion is the cause
of variations in the deuterium abundance.  In Figure \ref{tio}
we compare the abundances of Ti and O, which are both $\alpha$
elements whose enrichment comes mainly through Type II
supernovae of massive stars.  Applying the above logic,
one would expect a flat distribution of Ti/O vs. D/H in the infall
scenario, but a positive correlation with D/H for dust depletion.
The Spearman rank correlation coefficient is $-$0.1, which
indicates that there is no significant correlation between Ti/O
and D/H, and the negative value actually indicates the data
tend to an anti-correlation.  Unfortunately, this test
is not conclusive, since the very mild dependence of Ti/H
on D/H (Figure \ref{tih}) spans a factor of three less (0.5 dex)
than the scatter in the Ti/O ratios at a given D/H.  

We have already argued that a `standard' infall scenario with
low metallicity, enhanced D/H gas would cause an anti-correlation
between [X/H] and D/H.  Although D-poor infall could qualitatively
explain the positive correlations in
Figure \ref{tih}, again the gradients provide strong evidence
against this possibility.  If infalling gas has a low metallicity, but a
solar abundance pattern, then in a plot of [X/H] vs. D/H, every
element should have the same gradient (ignoring all other effects
such as depletion, ionization etc.).  This is simply due to the
fact that each element is diluted by the same amount and with a
logarithmic ordinate, the $\Delta y$ is a measure of the dilution
factor.  The very different slopes seen in Figure \ref{tih}
therefore argue against a D-poor infall model, although we can not
rule out more contrived models
with highly non-solar abundances.  However, combined with other
evidence against the infall model, such as the
relative constancy of certain elements such as O and Kr
(Oliveira et al. 2005; Cartledge et al. 2004) and the sharp transition
from approximately constant D/H inside the local bubble to
a large scatter at distances beyond 100 pc, which would require 
a coincidentally very localized region of infall (e.g.
de Avillez \& Mac Low 2002), we conclude that infall is 
unlikely to be the main cause of local D/H variations.  This is
not to say that the local ISM is a closed box, in fact Romano
et al. (2006) require some infall to explain the modest astration
between the primordial D abundance and the local values. 

\subsection{Scatter}\label{scatter_sec}

In closing this section, we note that for a given D/H, there is a wide range of
observed [Ti/H] abundances, spanning up to a factor of three
(Figure \ref{tih}).  The scatter means that some D/H$\sim$10 ppm sightlines
have [Ti/H] as high as the highest D/H sightlines, and vice versa.
Linsky et al. (2006) discussed a number of factors that might contribute
scatter in the relationship between element depletion and D/H
such as line saturation of DI transitions, ionization corrections
and N(HI) determination.  The former of these may
affect our results, although when saturation is obvious, we
remove the sight line from our analysis (e.g. HD 206773).  However,
we expect the latter two concerns to be alleviated by our
choice of high N(HI) sight lines.  The more obvious damping
wings should facilitate the determination of N(HI) that can
be more challenging for log N(HI) $<$ 19 and the gas will
be approaching complete self-shielding at these high column
densities.    
In short, the observed scatter in Ti/H is most likely to represent
true variations in the gas-phase abundance, not systematic effects.
Such intrinsic scatter may be expected due to the dynamic nature
of the ISM as dust is formed and ablated and the gas phase responds
to photoionization and even incorporation into molecules.  Steigman
et al. (2007) have also argued that some amount of infalling gas
that is not well mixed can not be ruled out.  This will also
contribute to intrinsic scatter in abundances.

\section{Conclusions}

We present 12 new measurements of Ti/H for local Galactic sightlines
for which D/H measurements exist, almost tripling the sample
from Prochaska et al. (2005).  We investigate elemental abundance
trends in the context of depletion of D onto grains as the cause for
the large variation seen in D/H in the local Galactic disk (e.g.
Wood et al. 2004 and references therein).  Our data support the tentative
correlation between Ti/H and D/H found by Prochaska et al. (2005),
showing a correlation at 97\% significance.  However,
it is not clear that this correlation can be explained by a simple depletion  
model.  This conclusion is based on 1) the much
shallower dependence of Ti/H on D/H than Si or Fe, despite the
fact that the former is much more refractory and 2) the
lack of a steep correlation between [Ti/H] and \avgnh.  If the variation in Ti
abundances of our sample sightlines were dominated by varying
dust depletion, we would expect a very steep correlation with
\avgnh, as has been previously found (Gondhalekar 1985; Jenkins 2004).
Moreover, the correlation of Si/H versus D/H is steeper than
between Fe/H and D/H, contrary to what is expected if these
relations trace depletion, since Fe is more refractory than Si.
This may mean that whatever effect is causing the flattening of the
Ti depletion correlations that are normally seen in the ISM
at large, may also affect Fe and Si, although probably to a lesser
degree.  However, the complex and dynamic nature of the ISM and
dust depletion no doubt complicates the above simplistic expectations.
Nonetheless, the shallow abundance correlations exhibited by
titanium are somewhat surprising given previous ISM studies, and it is
unclear why the depletion correlations have broken down in this
sample.  It may be due to superposition of the local bubble components
whose physical environment could be quite different from
the predominantly neutral disk ISM.
However, only using stars at distances $>$ 400 pc does not 
change our results and many stars in the Jenkins (2004) sample
are at distances less than 500 pc.  
Finally, we have also argued that the gradients
of [Si/H], [Fe/H] and [Ti/H] vs. D/H do not support a simple model of infall
of metal-poor gas with solar abundances as the dominant cause of
local D variations.  

\medskip

In summary, whilst the depletion of D onto grains remains a
plausible explanation for the variation in D/H in the local ISM, 
our results show that this interpretation is far from clear cut.

\section*{Acknowledgments}

We are very grateful to Cristina Oliveira who provided detailed comments
and expert advice on an earlier version of this manuscript 
and was generous in communicating results in advance of publication. 
We also benefitted from useful discussions and suggestions
with Chris Howk, Ed Jenkins, Jeff Linsky, Max Pettini and Blair Savage.
SLE is supported by an NSERC discovery grant and
SL was partly supported by the Chilean
{\sl Centro de Astrof\'\i sica} FONDAP No. 15010003, and by FONDECYT grant 
N$^{\rm o}1060823$.
Some of the data presented herein were
obtained at the W.M. Keck Observatory, which is operated as a
scientific partnership among the California Institute of Technology,
the University of California and the National Aeronautics and Space
Administration. The Observatory was made possible by the generous
financial support of the W.M. Keck Foundation.  The authors wish to
recognize and acknowledge the very significant cultural role and
reverence that the summit of Mauna Kea has always had within the
indigenous Hawaiian community.  We are most fortunate to have the
opportunity to conduct observations from this mountain.

\end{document}